\documentstyle[12pt]{article}
\advance\oddsidemargin	by -1.5in
\advance\evensidemargin by -1in
\advance\topmargin	by -0.2in
\newcommand\beq{\begin{equation}}
\newcommand\eeq{\end{equation}}
\newcommand\beqn{\begin{eqnarray}}
\newcommand\eeqn{\end{eqnarray}}

\def\itemitem{\par\indent \hangindent2\parindent \textindent}

\begin{document}

\baselineskip  14pt
\pagestyle{empty}

\begin{center}
{\large\bf High $\bf Q^2$ Probe of Nuclear Spectral Function\\
and Color Transparency}\\ \bigskip
	       BORIS Z. KOPELIOVICH\\
	       {\em Joint Institute for Nuclear Research,\\
	       Laboratory of Nuclear Problems,	\\
	       Head Post Office, P.O. Box 79,
	       10100 Moscow,
	       Russia}\\
e-mail:
boris@bethe.npl.washington.edu\\
\end{center}

\begin{center}
{\large Abstract}

\end{center}

\baselineskip  12pt
{\small
\rightskip=3pc
 \leftskip=3pc
Contrary to widespread opinion, color transparency (CT)
brings essential ambiguity to, rather than helps in, study of
the nuclear spectral function in quasielastic lepton scattering,
$A(l,l'p)A'$, at high $Q^2$.  Although the nuclear attenuation
vanishes, the final state interaction (FSI) of a small-size
ejectile wave packet, propagating through nuclear matter,
remains.  It manifests itself in a substantial, but uncertain,
longitudinal momentum transfer to the nuclear medium.  We predict
a strong Fermi-momentum bias of the nuclear transparency at high
$Q^2$, which enters as a factor at the nuclear spectral function,
and makes uncertain the results of measuring the
high-momentum tail of Fermi distribution.

\vspace{1.5cm}

\baselineskip 14pt

\rightskip=0pc
 \leftskip=0pc

According to expectations of PQCD the nuclear attenuation in
quasielastic electron scattering, $A(e,e'p)A'$, vanishes at high
$Q^2$, since the ejectile average size is small and it is
supposed
to demonstrate color transparency (CT) effects$^{1,2}$.  So one
may conclude naively that the ejectile
proton carries undisturbed information about the initial Fermi
momentum of the struck proton due to suppression of FSI by CT.
Thus one gets a perfect tool for study of the nuclear spectral
function.

The purpose of this talk is to demonstrate that CT, on
the contrary to above expectations, does not rule out the problem
of FSI. Moreover, it makes it much more difficult than at low
$Q^2$. Just CT breaks factorization of the hard quasielastic
scattering cross section, is responsible for an
longitudinal momentum transfer to the nucleus during FSI,
and causes an asymmetry of nuclear transparency on missing
momentum$^3$. These facts leave no hope to use a high-$Q^2$
quasielastic scattering of leptons and hadrons for a reliable
measurement of the nuclear spectral function.

Let us consider the imaginary part of the amplitude
corresponding to all intermediate particles being on mass shell.
To have CT different intermediate states have to cancel each
other in the imaginary part of the amplitude.  Elastic electron
scattering on a free proton target corresponds to a fixed Bjorken
variable, $x_B=1$, which is fixed only by electron momenta.  This
fixes the mass of the ejectile at proton mass.	We are arriving
at a puzzling conclusion that only proton, rather than a set of
different hadronic states, is produced.  Thus no cancellation is
possible, i.e.	no CT.

However in quantum mechanics one has to define how the size of
the ejectile is
measured, because the measurement it self is known to affect the
result. In order to minimize such an influence, one could put
a size detector (second scattering center) far apart from the
target proton. Then the above
conclusion is correct: due to the time evolution of the ejectile
only a proton will reach the distant detector.

To observe CT one has to put the size detector at a short
distance from the target proton.  Then the proton can not be
treated as at rest, according to the uncertainty principle.  Its
momentum is distributed with a width of the order of the inverse
distance from the detector (Fermi motion).  As a result, some
heavier states can be produced.  The closer the target is to the
size detector the more states are produced, the complete is the
cancellation.

Thus the Fermi motion of a bound nucleon is a source of CT in
quasielastic scattering$^{3,4}$.
On the other hand it restricts the amount of CT because the Fermi
motion spectrum in nuclei is concentrated mainly within momenta
of
about $k_F = 0.2$GeV/c. If the Bjorken variable, $x_B = Q^2/(2
m_p \nu)$, is fixed at $x_B=1$,
the electron knocks out a proton in the hard scattering on a
bound proton, only if the latter was at rest.
To produce an excited state of mass $m^*$ in the hard
scattering, the
target nucleon must have an initial momentum in the direction
opposite to the photon, $k_z \approx -(m^{*2} + m_p^2)/2\nu$. So
the mass spectrum of produced states is restricted by:

\begin{equation}
m^{*2} < m_p^2 + 2 \nu k_F
\label{2.6a}
\end{equation}

Consequently even if the ejectile in $(e,e'p)$ reaction has a very small size,
$\rho^2 \approx 1/Q^2$, the nucleus, as a quantum size detector, is insensitive
to such a small size. Only the hadronic states, which can be
created with available Fermi momenta contribute to the
produced wave packet.Therefore a nucleus  can resolve only size
larger than:

\begin{equation}
\rho^2 \geq \frac{m_p}{k_F}\frac{1}{Q^2}.
\label{2.6b}
\end{equation}

This restriction evidently reflects the fact that the size of the
ejectile cannot  be measured  after the hard
interaction at shorter distance than the mean internucleon
separation in nuclei.
The latter is just related to the mean Fermi momentum.
\smallskip

Let us consider the $x_B$-dependence of the typical observable,

\beq
Tr=\frac{\sigma^A}{A\;\sigma^N} ,
\label{1.3a}
\eeq
called {\it nuclear transparency} (or transmission coefficient),
because it naively looks that when one assumes factorization of
the nuclear cross section $\sigma^A$. Later on we will see that
it is not so.

Starting with $x_B=1$, note that the Fermi motion is used
ineffectively: only a half of the quasielastic
peak with $k_z<0$ is used, no states on the mass shell are
produced
at Fermi-momenta $k_z>0$. This situation is illustrated
schematically in fig.1a. Varying $x_B$, one can shift the mass
spectrum of hadronic states along the $k_z$ axis, increasing or
suppressing the interval of mass covered by the Fermi momenta.
Thus one can effectively change the transverse size of the
produced wave packet, and the handle to do it is the Bjorken
variable$^3$. For instance, if we decrease $x_B$ down to
$x_B\approx 1-k_F/m_p$, the mass squared interval, $\Delta
m^2$,
doubles in comparison with $x_B=1$. This is
illustrated in fig.1b, and follows from the
approximate relation between $x_B$, Fermi moment $\vec k$ and
the produced mass:

\beq
M^2\approx m_p^2+Q^2\;\frac{1-x_B}{x_B}-2qk_z ,
\label{2.6ba}
\eeq
where $\vec q$ if the 3-momentum of the photon.
  As a consequence of enlarged available mass interval, $Tr$
will rise. If one keeps decreasing $x_B$ one can arrive even at
nuclear
antishadowing, $Tr>1$. Indeed at some $x_B$ the lighest hadronic
state, proton, is pushed out of the Fermi distribution, i.e.
its direct production is extremely suppressed.
However it can be still effectively produced via some heavier
intermediate state. The nuclear transparency (\ref{1.3a}),
in the case of $(e,e'p)$ reaction should be normalized by the
probability to have Fermi momentum $k_z$, corresponding to the
proton production. Thus, at $x_B<1-k_F/m$ the Fermi-distribution
suppresses the denominator of (\ref{1.3a}), than the numerator.
Therefore $Tr$ increases and may cross unity. The latter depends
much on the form of the edge of the Fermi momentum distribution
and the mass spectrum of produced states. An explicit example is
given below.

One gets an opposite result increasing $x_B$. The larger is
$x_B$, the less hadronic states contribute to the ejectile wave
packet. At last at $x_B\geq 1+k_F/m$ all states except a proton
are pushed out of the Fermi distribution on the $k_F$ scale, as
is shown schematically in fig.1c. It means that in this case the
ejectile is simply a
proton, the Glauber approximation is exact (up to usual inelastic
corrections, which are as small as in the total $hA$ cross
section).

Thus we predict an $x_B$-asymmetry of nuclear transparency$^3$,
 which
is the direct reflection of the deep quantum mechanical origin of
the CT phenomenon.

After the Fermi bias effect
was claimed and estimated in Ref.3, the calculations were
repeated by
other authors$^{5,6}$. Nucleon correlations in nuclear density
matrix
were taken into account more carefully in Ref.6. This provides
a distortion of the effective Fermi momentum distribution by the
absorption of the recoil proton in nuclear matter. As a result
a weak $x_B$ dependence  of nuclear transparency proves to appear
even in the Glauber approximation. The form of the edge of Fermi
momentum distribution, and the high-momentum tail, are most
important for the magnitude of nuclear transparency at $x_B\leq
1-p_F/m_p$. Depending on it, the interplay
between the direct production of proton and production via
intermediate excitations, can result in a nuclear antishadowing,
like in Ref.3, or a shadowing, like in Ref.6. Unfortunately
the high-momentum tail of the Fermi distribution is poorly
known. We disagree with the statement
of Ref.6, that the lack of nuclear antishadowing, they got, is
due to the coherency constraint, since the latter was
completely included in the evolution operator of Ref.3.

Let us estimate roughly the scale of the effect. Nuclear
transparency (\ref{1.3a}) is read,

\begin{equation}
Tr=
\frac{\sum_{\alpha}\int
d^2b\left |\int^{\infty}_{-\infty}
dz\;e^{ik_zz}\psi^{\alpha}_A(b,z)
\langle p|\hat V(z,\infty)|i\rangle\right |^2}
{A\;W_A(k_z)\;|\langle p|i\rangle|^2} ,
\label{2.6c}
\end{equation}
where $\vec k$ is the "missing momentum" in the $(e,e'p)$
reaction, defined as $\vec k=\vec q-\vec p_p$.
$\Psi_A^\alpha$
is the nuclear wave function in the shell $\alpha$.
$W_A(k_z)=\int d^2k_T~W_A(\vec k)$, where $W_A(\vec k)$ is the
destribution function of Fermi momenta in the nucleus.
$\hat
V(z,z')\; exp(ik_z(z-z'))$ is an operator of evolution of the
ejectile wave packet in the nuclear medium,

\begin{equation}
i\frac{d}{dz} | P \rangle = \hat V | P \rangle ,
\label{2.3h}
\end{equation}

$|P\rangle$ denotes the set of states $|p\rangle$, a proton, and
$|p^*\rangle$, which is an effective state reproducing
contribution of all the diffractive excitations of the proton.
The center of gravity of these excited states is the mass, $m^*$,
of $|p\rangle$. The evolution operator $\hat V$ has the form,

\begin{equation} \hat V = \left(
\begin{array}{cc}
p -i \sigma^{hN}_{\rm tot}/2\;\rho_A(z) & \frac{\alpha}{\sqrt{1-
\alpha^2}} i\sigma^{hN}_{\rm tot} \rho_A(z)\\
\frac{\alpha}{\sqrt{1- \alpha^2}} i\sigma^{hN}_{\rm tot}
\rho_A(z) & p -q_{\rm hh^*} - i\sigma^{hN}_{\rm tot}/2\;\rho_A(z)
\end{array}\right)
\label{2.3i}
\end{equation}
where p is the proton momentum; $q_{\rm hh^*}$ is the
longitudinal momentum transfer,

\beq
q_{\rm hh^*}
=\frac{{m^*}^2-m^2}{2\;\nu}
\label{2.3d}
\eeq

The initial state $|i\rangle$ is assumed to be a noninteracting
one due to CT. This puts constraints on its decomposition,
$|i\rangle=\alpha\;|p\rangle + \sqrt{1-\alpha^2}\;|p^*\rangle$.
The parameter $\alpha$ is expressed in terms of the ratio of
the forward diffractive dissociation and of the elastic scattering cross
sections as:
\begin{equation}
\frac{\sigma_{\rm dd}}{\sigma_{\rm el}} =
\frac{\alpha^2}{1- \alpha^2}
\label{2.3g}
\end{equation}

Let us represent the nuclear density matrix in the form,

\beq
\rho(\vec r_1,\vec r_2)=
\sum_\alpha\Psi_A^\alpha(\vec r_1)\Psi_A^{\alpha *}(\vec r_2) ,
\label{2.6d}
\eeq
in the form

\beq
\rho(\vec r_1,\vec r_2)=\rho_A(\vec r)\;P_A(\vec\Delta),
\label{2.6e}
\eeq
where $\vec r=1/2(\vec r_1+\vec r_2), \vec\Delta=\vec r_2-\vec
r_1$.

The function $P_A(\vec\Delta)$ satisfies the following
conditions,

\beq
P_A(0)=1
\label{2.6f}
\eeq
\beq
\int d^2\Delta\;P_A(\vec\Delta)\;e^{i\vec k\vec\Delta}=W_A(\vec
k) \label{2.6g}
\eeq

Using this, the expression (\ref{2.6c} is transformed to the
form,

\beqn
\lefteqn{Tr=
\left [A\;W_A(k_z)\;|\langle p|i\rangle|^2\right ]^{-1}
\int d^2b\int^{\infty}_{-\infty}
dz\;\rho_A(\vec b,z)}
\nonumber\\
&&\int^{\infty}_{-\infty}
d\Delta_z\;
e^{ik_z\Delta_z}\;P_A(\Delta_z)\;
\langle p|\hat V(z-\Delta_z/2,\infty)|i\rangle\;
\langle p|\hat V(z+\Delta_z/2,\infty)|i\rangle^*
\label{2.6h}
\eeqn
Using equation (\ref{2.3h}) we can write a
recurrent relation,

\beq
\hat V(z\pm\Delta z/2,\infty) = \hat V(z\pm\Delta z/2,z)\hat
V(z,\infty)
\eeq

It follows from (\ref{2.6g}), that the
distribution $P_A(\Delta_z)$ is narrow,
$\langle\Delta_z^2\rangle^{1/2}\approx
\frac{\sqrt 6}{ k_F}\approx 2~fm$.
So the function $\hat V(z\pm\Delta_z/2,z)$ may be expanded over
this small parameter.

 \beq
\hat
V(z,z\pm\Delta_z/2)=\hat I(1\pm ip\Delta_z/2) \mp {i\over
2}~\Delta_z~\hat U + O(\Delta_z^2) ,
\label{2.6i}
\eeq
where $\hat I$ is a unit matrix.
We neglect the terms of the order of
$\langle\Delta_z^2\rangle$ and higher, using
smallness of the correlation length.
The conditions for such an approximation are:

\beq
{1\over 2}~\Delta z\Delta p\ll 1
\label{2.6j}
\eeq
\beq
{1\over 2}\sigma_{tot}\rho_A(\vec r)\Delta z\ll 1
\label{2.6k}
\eeq
The former is valid at high energies, starting from a few GeV. The
latter depends on the nuclear density at the point of
interaction. It is satisfied even at the very center
of a nucleus, but becomes more exact at the nuclear edge,
which is the most
important for the process, if it is far from the saturation of
CT.

Note that the terms of the order of $\langle\Delta_z^2\rangle$
give rise to a weak $k_z$-dependence of $Tr$ even in Glauber
approximation (compare with Ref.6).

When (\ref{2.6i}) is substituted into (\ref{2.6h}), the imaginary
part of $\hat U$
cancels. The real part of $\hat U$ gives an additional phase factor,
$exp(i~\Delta p\Delta z/2)$ to each amplitude $\langle p|\hat
V(z,\infty)|p^*\rangle$. This shift of phase leads to the shift
of the
argument of $W_A(k)$ after integration over $\Delta z$. To take it
into account let us introduce a shifted state

\beq
|\tilde i\rangle = \alpha |p\rangle +
\sqrt{1-\alpha^2}\;\sqrt{\frac{W_A(k+\Delta p)}
{W_A(k)}}\;|p^*\rangle
\label{2.6l}
\eeq

Such a transformation of the initial state is an approximation
(see Ref.6), which is numerically quite precise.

Finally, (\ref{2.6h}) is transformed into

\beq
Tr(x_B) = \frac{\int d^2b\;\int_{-\infty}^{\infty}dz\;\rho_A(b,z)\;
|\langle p|\hat V(z,\infty)|\tilde i\rangle |^2}{A\;|\langle
p|i\rangle|^2}
\label{2.6m}
\eeq

To evaluate  we use the
two-channel approximation$^7$. We are not
aimed here to provide reliable numerical predictions, but only
raise principal questions, and need only to get a rough estimate
of
the effect. For this reason we use the simple two-channel
approximation to evaluate $\hat V(z,\infty)|\tilde i\rangle$,
and a simplified form of Fermi momentum distribution with
a Gaussian parameterization,

\begin{equation}
 W_A(k) = \frac{3}{2\pi k_F^2}\exp(-3k^2/2k_F^2).
\label{2.6n}
\end{equation}

The results are shown in fig.2	as a function of
$x_B$ for $Q^2$ equal to 7 GeV$^2$, 15 GeV$^2$ and 30 GeV$^2$.
We use $m^*=1.6\;GeV$ and $\alpha=0.1$.
As mentioned previously, at $x_B > 1$ the nuclear transparency is
small and close to the expectation of the Glauber model.

At $x_B < 1$, on the contrary excited states are preferentially
produced; to the extent that the transparency even becomes larger
then 1 for small $x_B$.

The above calculations have a demonstrative character.
Nevertheless they are quite indicative of the
results that would be obtained with a better calculation.

We see that the $(l,l'p)$ reaction at  high $Q^2$ is a poor probe
of the nuclear spectral function.

The same concerns apply to wide-angle quasi-elastic
$(p,2p)$ scattering (see next section).

To conclude, a few comments in order.
\begin{itemize}

\item
CT in quasielastic scattering on nuclei is possible only due to
Fermi motion. However the finiteness of Fermi momenta strongly
violates the CT sum rule. This depends on Bjorken $x_B$ (or
missing momentum), which allows to handle the amount of CT. The
predicted $x_B$-asymmetry of nuclear transparency is a new
observable of CT, reflecting its a deep quantum mechanical
nature.

\item
The finiteness of available Fermi momenta and energy conservation
cut off most of hadronic states, which contribute to the CT sum
rule.  One might arrive at a wrong conclusion, forgetting about
it.  An example is the proposal$^{21}$ for CEBAF to study CT
effects in $(e,e'2p)$ on light nuclei.  Even at low $Q^2\approx
4\;GeV^2$ the authors predict about $50\%$ deviation from Glauber
model.	They use three channel model with masses
$m=m_p,\;m^*=1.6\;GeV$ and $m^{**}=3\;GeV$.  It is easy to check
however, that two latter states are too heavy to be produced with
reasonable Fermi momentum at $Q^2=4\;GeV^2$.  Thus this model,
corrected for the conservation of energy, predicts no deviation
from the Glauber model in the CEBAF range of $Q^2$.

\item
Although different hadronic states produced in the hard
interaction use different Fermi momenta, the final
proton momentum is fixed.  The difference between the
longitudinal momenta
of hadronic states after the hard interaction, is compensated by
longitudinal momentum
transfer in diffractive FSI.  Thus a small size ejectile,
consisted of many hadronic waves, transfer to the nuclear matter
a positive (relative to the photon direction) longitudinal
momentum of the order of mean Fermi momentum.  This keeps being
valid at any $Q^2$.  Therefore CT does not mean an overall
suppression of FSI, but only strong cancellation in the
transmission amplitude.

\item
According to the wide spread opinion, the high $Q^2$ quasielastic
scattering, provides a clean tool for measurement of the nuclear
spectral function, because CT suppresses FSI.  Such a believe is
based on the classical treatment of CT.  It is amazing that a
correct quantum mechanical analyses leads to the opposite
conclusion:  CT does not help, but just spoils any opportunity to
measure the nuclear spectral function.	Firstly, according to the
previous comment, CT does mean that FSI exists, provided an
additional longitudinal momentum transfer, escaping detection.
It is unavoidable property of CT.  Secondly, if one
wants to measure the large momentum tail of Fermi distribution he
faces the Fermi bias of nuclear transparency, which cannot be
calculated reliably.  One is safe of these problems only at $x_B
< 1-p_F/m_p$, where the Glauber approximation is correct.  One
may wonder however in this case, why does he need high $Q^2$ at
all.

\item
The observed phenomenon of Fermi bias sheds light$^{15}$ on
the puzzling results of measurement of nuclear transparency in
quasielastic $(p,2p)$ scattering in the BNL experiment$^7$.
Actually the data were distributed over missing momentum, so the
asymmetry predicted here for such a distribution has a direct
concern to interpretation of the data.
Calculations were performed in Ref.8 within the same
two-component
model, as used in Ref.3.  It is found that the missing momentum
asymmetry nicely explains the observed drop of transparency at
$12\;GeV/c$ beam momentum, which have just confused theorists,
expecting a rising energy dependence.

\item Note that it follow from the above consideration of
quantum effects, that CT effects are essentially due to the
observation of the ejectile proton in the final state. This is
why no CT effect is expected in inclusive deeply inelastic
scattering of leptons on nuclei.

\item
The finiteness of available Fermi momenta and energy conservation
cut off most of hadronic states, which contribute to the CT sum
rule.  One might arrive at a wrong conclusion, forgetting about
it.  An example is the proposal$^{9}$ for CEBAF to study CT
effects in $(e,e'2p)$ on light nuclei.  Even at low $Q^2\approx
4\;GeV^2$ the authors predict about $50\%$ deviation from Glauber
model.	They use three channel model with masses
$m=m_p,\;m^*=1.6\;GeV$ and $m^{**}=3\;GeV$.  It is easy to check
however, that two latter states are too heavy to be produced with
reasonable Fermi momentum at $Q^2=4\;GeV^2$.  Thus this model,
corrected for the conservation of energy, predicts no deviation
from the Glauber model in the CEBAF range of $Q^2$.

\end{itemize}

\smallskip

{\bf References}

\vspace{0.3cm}

\itemitem{1.} A.B.~Zamolodchikov, B.Z.~Kopeliovich and
L.I.~Lapidus, {\it JETP Lett.}	{\bf 33}, (1981) 595

\itemitem{2.} G.~Bertsch, S.J.~Brodsky, A.S.~Goldhaber and
J.G.~Gunion, {\it Phys.  Rev.  Lett.}  {\bf 47}, (1981) 297

\itemitem{3.} B.~Jennings ans B.~Kopeliovich, {\it Phys. Rev.
Lett.} {\bf 70}, (1993)3384

\itemitem{4.} A.~Biamconi, S.~Boffi, D.E.~Kharzeev, preprint
FNT/T-92/38, (1992)

\itemitem{5.} A.~Biamconi, S.~Boffi, D.E.~Kharzeev,
FTN/T-92/47,

\itemitem{6.} N.N.~Nikolaev et al., J\"ulich preprint
KFA-IKP(Th)-1993-16(1992)

\itemitem{7.} B.Z.~Kopeliovich and L.I.~Lapidus, {\it JETP
Lett.} {\bf 32}, (1980)595

\itemitem{8.} B.~Jennings and B.~Kopeliovich, PANIC XIII, Book
of Abstracts, p.165

\itemitem{9.} K.~Egiyan et al., PANIC XIII, Book of Abstracts,
p.178, 1993

\vspace{1cm} \

{\bf Figure captions}
\vspace{0.3cm} \\

{\bf Fig.1} \\
Schematic picture, showing the distribution  of exited proton
states over the Fermi momentum, versus Feynman variable $x_B$.

{\bf Fig.2} \\
The nuclear transparency in $(e,e'p)$ as function of Bjorken
variable $x_F$, versus $Q^2=7,\;15$ and $30\;GeV^2$

\end{document}